\documentclass[12pt]{article}

\usepackage{amsmath}
\usepackage{amssymb}
\usepackage{amsfonts}
\usepackage{latexsym}
\usepackage{color}

\catcode `\@=11 \@addtoreset{equation}{section}

\catcode `\@=12

\newcommand{\be}{\begin{equation}}
\newcommand{\en}{\end{equation}}
\newcommand{\bea}{\begin{eqnarray}}
\newcommand{\ena}{\end{eqnarray}}
\newcommand{\beano}{\begin{eqnarray*}}
\newcommand{\enano}{\end{eqnarray*}}
\newcommand{\bee}{\begin{enumerate}}
\newcommand{\ene}{\end{enumerate}}

\newcommand{\M}{\mathfrak M}

\newcommand{\A}{{\mathfrak A}}

\newcommand{\R}{\cal R}
\newcommand{\mc}{\mathcal}

\newcommand{\Sc}{{\cal S}}

\newcommand{\1}{1 \!\! 1}

\newcommand{\Hil}{\mc H}

\catcode `\@=11 \@addtoreset{equation}{section}
\catcode `\@=12

\begin{document}

\thispagestyle{empty}

\vspace*{2cm}

\begin{center}
{\Large \bf  Few simple rules to fix the  dynamics of classical systems using operators}\\[10mm]

{\large F. Bagarello}\\
  DIIETCAM,
Facolt\`a di Ingegneria,\\ Universit\`a di Palermo, I-90128  Palermo, Italy\\
e-mail: fabio.bagarello@unipa.it\\ Home page:
www.unipa.it/fabio.bagarello\\

\vspace{3mm}

\end{center}

\vspace*{2cm}

\begin{abstract}
\noindent We show how to use operators in the description of {\em exchanging processes} often taking place in (complex) classical systems. In
particular, we propose a set of rules giving rise to an {\em hamiltonian} operator for such a system $\Sc$, which can be used to deduce the
dynamics of $\Sc$

\end{abstract}

\vspace{2cm}


\vfill


\newpage

\section{Introduction and motivations}

In a series of recent papers we have used an operatorial approach in the description of classical systems, with few or with many degrees of
freedom, \cite{bag1}-\cite{ff3}. In particular, we have shown how canonical commutation and anticommutation relations (CCR and CAR
respectively) can be used in the analysis of simplified stock markets, as well as in the description of simpler dynamical systems, like those
arising from {\em love affairs}. We have also adopted the same general settings in the analysis of migration processes and of population
dynamics.

The main ingredient in our approach is the {\em hamiltonian} operator $H$ of the system $\Sc$ we are interested in, which is used to deduce the
time evolution of $\Sc$, see below. This paper is devoted to discuss a minimal set of rules which should be adopted to write down $H$. Some
examples of hamiltonians found this way will be discussed. However, the dynamical content of these hamiltonians will not be considered here,
since it was already discussed elsewhere, \cite{bag1}-\cite{ff3}.

\vspace{2mm}

The paper is organized as follows: in the rest of this section we review few known fact on CCR. We will not discuss here CAR since they will
only play a minor role in Section III.

In  Section II we propose our  set of rules useful to determine the analytic expression of the hamiltonian of a system $\Sc$.

In Section III we discuss few examples, while Section IV contains our conclusions.

\vspace{3mm}

The reason why the operator $H$ assumes a crucial role in our approach is because the dynamical behavior of $\Sc$ is here assumed to be given
by the Heisenberg equation of motion:

let $\Hil$ be an Hilbert space and $B(\Hil)$ the set of all the bounded operators on $\Hil$.   Let $\Sc$ be our physical system and $\A$ the
set of all the operators useful for a complete description of $\Sc$, which includes the {\em observables } of $\Sc$.  The time evolution of
$\Sc$ is assumed to be given by the self-adjoint hamiltonian $H=H^\dagger$  of $\Sc$, which in standard quantum mechanics represents the energy
of $\Sc$. In the { Heisenberg} picture the time evolution of an observable $X\in\A$ is expressed by \be X(t)=e^{iHt}Xe^{-iHt}\label{11}\en or,
equivalently, by the solution of the differential equation \be \frac{dX(t)}{dt}=ie^{iHt}[H,X]e^{-iHt}=i[H,X(t)],\label{12}\en where
$[A,B]:=AB-BA$ is the {\em commutator } between $A$ and $B$. The time evolution defined in this way is usually a one parameter group of
automorphisms of $B(\Hil)$. It might seem that adopting the {Heisenberg} picture in the description of classical systems may appear
unappropriate. However, as discussed in our previous literature as well as in many other papers on similar subjects, see for instance
\cite{baa}-\cite{tulu}, this approach is justified {\em a posteriori} since, at least for simple systems, it produces exactly that time
evolution which one expects to find. We should also mention that  the uncertainty principle arising from the non abelianity of the operators
involved in the description of $\Sc$, does not appear in our approach, since all the observables of $\A$ do commute. Other authors, on the
other hand, because of what they are interested in, consider such an uncertainty a richness and not a problem of a quantum view to complex
systems, \cite{seg}.

In our approach a special role is played by the so called CCR: we say that a set of operators $\{a_l,\,a_l^\dagger, l=1,2,\ldots,L\}$ satisfy
the CCR if the following hold:\be [a_l,a_n^\dagger]=\delta_{ln}\1,\hspace{8mm} [a_l,a_n]=[a_l^\dagger,a_n^\dagger]=0, \label{a3}\en for all
$l,n=1,2,\ldots,L$. Here $\1$ is the identity operator on $\Hil$. These operators, which are widely analyzed in any textbook in quantum
mechanics, see \cite{mer} for instance, are those which are used to describe $L$ different {\em modes} of bosons. From these operators we can
construct $\hat n_l=a_l^\dagger a_l$ and $\hat N=\sum_{l=1}^L \hat n_l$ which are both self-adjoint. In particular $\hat n_l$ is the {\em
number operator } for the l-th mode, while $\hat N$ is the {\em number operator of $\Sc$}.

The Hilbert space of our system is constructed as follows: we introduce the {\em vacuum} of the theory, that is a vector
$\varphi_{0,0,\ldots,0}$ which is annihilated by all the operators $a_l$: $a_l\varphi_{0,0,\ldots,0}=0$ for all $l=1,2,\ldots,L$. Then we act
on $\varphi_{0,0,\ldots,0}$ with the  operators $a_l^\dagger$ and their powers: \be
\varphi_{n_1,n_2,\ldots,n_L}:=\frac{1}{\sqrt{n_1!\,n_2!\ldots n_L!}}(a_1^\dagger)^{n_1}(a_2^\dagger)^{n_2}\cdots
(a_L^\dagger)^{n_L}\varphi_{0,0,\ldots,0}, \label{14}\en $n_l=0,1,2,\ldots$ for all $l$. These vectors form an orthonormal set and are
eigenstates of both $\hat n_l$ and $\hat N$: $\hat n_l\varphi_{n_1,n_2,\ldots,n_L}=n_l\varphi_{n_1,n_2,\ldots,n_L}$ and $\hat
N\varphi_{n_1,n_2,\ldots,n_L}=N\varphi_{n_1,n_2,\ldots,n_L}$, where $N=\sum_{l=1}^Ln_l$. Moreover using the  CCR we deduce that \be\hat
n_l\left(a_l\varphi_{n_1,n_2,\ldots,n_L}\right)=(n_l-1)(a_l\varphi_{n_1,n_2,\ldots,n_L})\label{1ag1}\en and \be\hat
n_l\left(a_l^\dagger\varphi_{n_1,n_2,\ldots,n_L}\right)=(n_l+1)(a_l^\dagger\varphi_{n_1,n_2,\ldots,n_L}),\label{1ag2}\en for all $l$. For these
reasons the following interpretation is given: if the $L$ different modes of bosons of $\Sc$ are described by the vector
$\varphi_{n_1,n_2,\ldots,n_L}$, this implies that $n_1$ bosons are in the first mode, $n_2$ in the second mode, and so on. The operator $\hat
n_l$ acts on $\varphi_{n_1,n_2,\ldots,n_L}$ and returns $n_l$, which is exactly the number of bosons in the l-th mode. The operator $\hat N$
counts the total number of bosons. Moreover, the operator $a_l$ destroys a boson in the l-th mode, while $a_l^\dagger$ creates a boson in the
same mode. This is why $a_l$ and $a_l^\dagger$ are usually called the {\em annihilation} and {\em creation} operators.

The Hilbert space $\Hil$ is obtained by taking the closure of the linear
span of all these vectors.

A similar construction can be repeated starting with CAR, but we will not consider this possibility here since it is not essential for the
general analysis we will discuss in this paper.

 \vspace{2mm}

An operator $Z\in\A$ is a {\em constant of motion} if it commutes with $H$. Indeed in this case equation (\ref{12}) implies that $\dot Z(t)=0$,
so that $Z(t)=Z$ for all $t$.

The vector $\varphi_{n_1,n_2,\ldots,n_L}$ in (\ref{14}) defines a {\em vector (or number) state } over the algebra $\A$  as
\be\omega_{n_1,n_2,\ldots,n_L}(X)= \left<\varphi_{n_1,n_2,\ldots,n_L},X\varphi_{n_1,n_2,\ldots,n_L}\right>,\label{15}\en where
$\left<\,,\,\right>$ is the scalar product in  $\Hil$. As we have discussed in \cite{bag1}-\cite{ff3}, these states are used to {\em project}
from quantum to classical dynamics and to fix the initial conditions of the system.

\section{The rules}

As already discussed in the Introduction, the main interest in this paper is to produce a sort of recipe which has to be used to write down the
hamiltonian $H$ of the classical system $\Sc$ we are interested in. To simplify our analysis, let us first suppose that $\Sc$ consists of two
main interacting parts, $\Sc_1$ and $\Sc_2$, the {\em actors} of the game, whose union reproduces $\Sc$ and which have no intersection:
$\Sc=\Sc_1\cup\Sc_2$ and $\Sc_1\cap\Sc_2=\emptyset$. Suppose now that $\Sc_1$ and $\Sc_2$ can exchange something, $\M$, which can only take
integer values\footnote{We could relax this assumption by assuming that the values of $\M$ are discrete rather than integer.}. A typical
example of this situation is in stock markets, where two traders $\Sc_1$ and $\Sc_2$, exchange money (and shares). Other examples are discussed
in \cite{ff2} and \cite{bag5}, where what is exchanged is {\em mutual affection} (in other words, love!) between the two lovers. In \cite{ff3}
we have two populations in different regions of a two-dimensional lattice, and they {\em exchange people}, i.e. there is people moving from one
region to the other.

Let us now introduce two annihilation operators, $a_1$ and $a_2$, related respectively to  $\M_1$ and $\M_2$,  and their conjugate creation
operators  $a_1^\dagger$ and $a_2^\dagger$. Here $\M_1$ is that part of $\M$ which {\em belongs} to $\Sc_1$: the money of the first trader, or
the number of shares in his portfolio, or jet the amount of love that Bob, the first lover, experiences for Alice, the second one, and so on.
As in the Introduction, these operators obey the following CCR: $[a_i,a_j] = [a_i^\dagger,a_j^\dagger]=0$,
$[a_i,a_j^\dagger]=\delta_{i,j}\,\1$. Calling $\varphi_{0,0}$ the vacuum of $a_1$, $a_2$, that is that vector of $\Hil$ annihilated by $a_1$
and $a_2$, $a_1\varphi_{0,0}=a_2\varphi_{0,0}=0$, the vector
$\varphi_{n_1,n_2}:=\frac{1}{\sqrt{n_1!\,n_2!\,}}\,{a_1^\dagger}^{n_1}{a_2^\dagger}^{n_2}\varphi_{0,0}$ describes a situation in which the
value of $\M_1$ is $n_1$ and  that of $\M_2$ is $n_2$; indeed, calling $\hat n_j:=a_j^\dagger a_j$ the related number operators, we know that
$\hat n_j\varphi_{n_1,n_2}=n_j\,\varphi_{n_1,n_2}$, $j=1,2$. In all the papers written so far, \cite{bag1}-\cite{ff3}, the eigenvalues of $\hat
n_j$, $n_j$, are considered directly related to the value of $\M_j$. Now we are ready to state the first rule of our construction:

\vspace{2mm}

{\bf Rule 1:--}{\em The exchange of $\M$ between $\Sc_1$ and $\Sc_2$ is modeled adding to the hamiltonian of $\Sc$ a term
$a_1^\dagger\,a_2+a_2^\dagger\,a_1$. If, for some reason, the model should be non-linear, then this contribution must be replaced by
${a_1^\dagger}^M\,a_2+a_2^\dagger\,a_1^M$, $M>1$ being a measure of the non linearity.}

\vspace{2mm}

The motivation of this rule is given by the action of $a_1^\dagger\,a_2+a_2^\dagger\,a_1$ on the vector $\varphi_{n_1,n_2}$:
$$
\left(a_1^\dagger\,a_2+a_2^\dagger\,a_1\right)\varphi_{n_1,n_2}\simeq\varphi_{n_1+1,n_2-1}+\varphi_{n_1-1,n_2+1},
$$
where the normalization constants are missing since they are not interesting for us, here. As we can see, what we get is a combination of two
vectors: the first one, $\varphi_{n_1+1,n_2-1}$, shows that the value of $\M_1$ is increased by one unit  while, simultaneously, $\M_2$
decreases by a unit. In the second contribution, $\varphi_{n_1-1,n_2+1}$, the opposite happens. In both cases, what it is going on is that
$\Sc_1$ and $\Sc_2$ are exchanging one unit of $\M$. Analogously, acting with ${a_1^\dagger}^M\,a_2+a_2^\dagger\,a_1^M$ on $\varphi_{n_1,n_2}$
would produce a combination of vectors $\varphi_{n_1+M,n_2-1}$ and $\varphi_{n_1-M,n_2+1}$, which is useful to introduce a possible asymmetry
between $\Sc_1$ and $\Sc_2$, or, from a dynamical point of view, a non-linearity in the dynamics of $\Sc$, \cite{ff2}.

One may argue why not to add simply $a_1^\dagger\,a_2$ in the hamiltonian. The reason is the following: if we don't consider both
$a_1^\dagger\,a_2$ and $a_2^\dagger\,a_1$, the final hamiltonian would not be self-adjoint, and this would create a lot of difficulties in
finding a reversible time evolution: for instance, if $H\neq H^\dagger$ then, among other problems, the norm of $e^{iHt}Xe^{-iHt}$ is different
from that of $X$, so that the probabilistic interpretation of the wave-function typical of quantum mechanics would be lost. This is obviously
related to the decay effects which we don't want to discuss here. We will briefly come back on this aspect in Section IV.

Let us now go the the second rule of our construction:

\vspace{2mm}

{\bf Rule 2:--} {\em The hamiltonian $H$ for $\Sc$ must contain a term, $H_0$,  such that, in absence of interaction between $\Sc_1$ and
$\Sc_2$, their related number operators, $\hat n_1$ and $\hat n_2$, stay constant in time.}

\vspace{2mm}

This is quite a natural assumption: if $\Sc_1$ and $\Sc_2$ do not interact, there is no reason for them to modify their situation, and in
particular there is no reason (and no possibility!) for exchanging units of $\M$. To be concrete, this means that, if at $t=0$ $\Sc$ is
described by the state $\varphi_{n_1,n_2}$, and if no interaction is contained in the hamiltonian, $H=H_0$, then at $t>0$ the system is still
described by $\varphi_{n_1,n_2}$ (but, at most, for an overall phase). It should be stressed that, however, this does not imply that in this
case there is no dynamics at all! What we are claiming is that $\hat n_j(t)=\hat n_j(0)$, but this does not imply that, for instance we also
necessarily have $a_j(t)=a_j(0)$. On the contrary, in many examples this is not so, see \cite{bag1} for such an example.

There is still another rule which is quite useful in the determination of $H$. For that it may be convenient to recall the notion of {\em
closed system}:  a system $\Sc$ is called closed if it has no interaction with the environment $\R$.

\vspace{2mm}

{\bf Rule 3:--} {\em If $\Sc$ is a closed system, the hamiltonian $H$ of $\Sc$ must commute with those global number-like operators related to
the observables which are not exchanged between $\Sc$ and $\R$.}

\vspace{2mm}

The motivation is, again, rather natural: as we have seen in the Introduction, all the observables which commute with the hamiltonian are {\em
integrals of motion}, so that they do not change with time. This is exactly what is expected to the {\em global quantities} of the system
$\Sc$, since they are not moving outside $\Sc$. A simple example of this situation is provided by the total number of shares of a certain type
in a closed market where the shares are not created or destroyed: if at $t=0$ this number is $N=n_1+n_2$, where $n_j$ is the number of shares
of that kind which belong to $\Sc_j$, then $N$ does not change with time, even if the number of shares in each trader's portfolio does change,
in general. In this case, {\bf Rule 3} reads $[H,\hat n_1+\hat n_2]=0$, while $[H,\hat n_1]\neq0$ and $[H,\hat n_2]\neq0$, in general.

\section{Examples}

In this section we will show how the rules described so far can be explicitly used in the analysis of some classical systems, and which kind of
hamiltonian are deduced.

\subsection{First example: love affair}

The first model we have in mind consists of a couple of lovers, Bob  and Alice, which mutually \emph{interact} exhibiting a certain
\emph{interest} for each other. Of course, there are several degrees of possible interest, and to a given Bob's interest  for Alice (LoA,
\emph{level of attraction}) there corresponds a related reaction (\emph{i.e.}, a different LoA) of Alice for Bob. In our previous decomposition
of $\Sc$ in $\Sc_1$ and $\Sc_2$, here Bob plays the role of $\Sc_1$, while Alice that of  $\Sc_2$, and $\M$ is the mutual affection between the
two. The bosonic operators associated to Bob are $a_1$, $a_1^\dagger$ and $\hat n_1=a_1^\dagger a_1$, while those associated to Alice are
$a_2$, $a_2^\dagger$ and $\hat n_2=a_2^\dagger a_2$. The (integer) eigenvalue  $n_1$ of $\hat n_1$ measures the value of the LoA that Bob
experiences for Alice: the higher the value of $n_1$ the more Bob desires Alice. For instance, if $n_1=0$, Bob just does not care about Alice.
We use $n_2$, the eigenvalue of $\hat n_2$, to measure the attraction of Alice for Bob. The {\em law of attraction} we have in mind states
that, if $n_1$ increases, then $n_2$ decreases and viceversa. This suggests to use the following self--adjoint operator to describe the
interaction between Alice and Bob: \be H=\lambda\left(a_1^{M}{a_2^\dagger}+a_2\,{a_1^\dagger}^M\right), \label{31}\en where $M$ describes a
sort of {\em relative behavior}, \cite{ff2}. This choice is written following {\bf Rule 1} of the previous section and it trivially satisfies
{\bf Rule 2}: if $\lambda=0$ there is no dynamics at all since $H=0$ and, as a consequence, $[H,\hat n_1] = [H, \hat n_2]=0$. Concerning {\bf
Rule 3}, it is an easy exercise to check that $I(t):=\hat n_1(t)+M\, \hat n_2(t)$ is a constant of motion: $I(t)=I(0)=\hat n_1(0)+M\, \hat
n_2(0)$, for all $t\in{\Bbb R}$, since $[H,I]=0$. Therefore, during the time evolution, a certain {\em global attraction} is preserved and it
can only be exchanged between Alice and Bob: notice that this reproduces our original point of view on the love relation between Alice and Bob:
the more Bob falls in love with Alice, the less Alice cares about Bob! If $M$ is fixed to be one then Bob and Alice react in the same way and
the model becomes exactly solvable, \cite{ff2}.

\vspace{3mm}

In \cite{ff2} we have also considered a love affair involving, other than Alice and Bob, a third actress, Carla, also having a relation with Bob. Our assumptions are the following: (1) Bob can interact with both Alice and Carla, but Alice
(respectively, Carla) does not suspect of Carla's (respectively, Alice's) role in Bob's life; (2) if Bob's LoA for Alice increases then Alice's LoA for Bob decreases and viceversa; (3) analogously, if Bob's LoA for Carla increases then Carla's
LoA for Bob decreases and viceversa; (4) if Bob's LoA for Alice increases then his LoA for Carla
decreases (not necessarily by the same amount) and viceversa.

Introducing now the operators $a_3$, $a_3^\dagger$ and $\hat n_3=a_3^\dagger a_3$ for Carla, and splitting the operators related to Bob in
two (i.e. $a_{12}$ and $a_{13}$ to describe  the interaction between Bob and, respectively, Alice and Carla) the
 hamiltonian which describes all these effects is the following:
\be H=\lambda_{12}\left(a_{12}^\dagger\,a_2+a_{12}\,a_2^\dagger\right)+\lambda_{13}\left(a_{13}^\dagger\,a_3+a_{13}\,a_3^\dagger\right)+
\lambda_{1}\left(a_{12}^\dagger\,a_{13}+a_{12}\,a_{13}^\dagger\right), \label{32} \en for some real values of $\lambda$'s. It can be easily
seen that the first contribution, $\lambda_{12}\left(a_{12}^\dagger\,a_2+a_{12}\,a_2^\dagger\right)$, describes the mechanism (2) above, while
$\lambda_{13}\left(a_{13}^\dagger\,a_3+a_{13}\,a_3^\dagger\right)$ is related to point (3). Point (4) is implemented by
$\lambda_{1}\left(a_{12}^\dagger\,a_{13}+a_{12}\,a_{13}^\dagger\right)$. These three contributions all trivially satisfy {\bf Rule 1} and,
again, {\bf Rule 2} is also verified: no interaction means that all the $\lambda$'s are zero, so that $H$ reduces to the zero operator, and all
the observables stay constant in time.  Let us  now introduce $\hat n_{12}=a^\dagger_{12}\,a_{12}$, describing  Bob's LoA for Alice, $\hat
n_{13}=a^\dagger_{13}\,a_{13}$, describing Bob's LoA  for Carla, $\hat n_{2}=a^\dagger_{2}\,a_{2}$, describing Alice's LoA  for Bob and $\hat
n_{3}=a^\dagger_{3}\,a_{3}$, describing Carla's LoA for Bob. If we define $J:=\hat n_{12}+\hat n_{13}+\hat n_2+\hat n_3$, which represents the
global level of LoA of the triangle, this is a conserved quantity: $J(t)=J(0)$, since $[H,J]=0$: no exhange with the environment is possible,
here! It is also possible to check that $[H,\hat n_{12}+\hat n_{13}]\neq 0$, so that the total Bob's LoA is not conserved during the time
evolution.

More details, as the equations of motion arising from these hamiltonians, their solutions and more extensions can be found in \cite{ff2} and
\cite{bag5}.

\subsection{Second example: competition between species and migration}

In this example we consider a two-dimensional region $\R$ in which two populations $\Sc_1$ and $\Sc_2$ are distributed. In \cite{ff3} we have
considered  these species as predators and preys, or as two migrant populations, moving from one part of $\R$ to another. Following the above
rules we can construct the hamiltonian of the full system $\Sc$. However, for reasons discussed in \cite{ff3}, it is convenient to use here
annihilation and creation operators satisfying CAR rather than CCR. This choice is motivated by a first technical and a second more substantial
reason: the technical reason is that we get finite dimensional Hilbert space for $\Sc$, while the more substantial reason is that, using CAR,
we automatically incorporate an upper bound for the densities of the two populations, which is a natural requirement for our {\em biological}
interpretation.

The starting point is the ({\em e.g.}, rectangular or square) region $\R$, which we divide in $N$ cells, labeled by $\alpha=1,2,\ldots,N$. In
each cell $\alpha$ the two populations, whose related operators are $a_\alpha$, $a_\alpha^\dagger$ and $\hat n^{(a)}_\alpha=a_\alpha^\dagger
a_\alpha$ for what concerns $\Sc_1$, and $b_\alpha$, $b_\alpha^\dagger$ and $\hat n^{(b)}_\alpha=b_\alpha^\dagger b_\alpha$ for $\Sc_2$, are
described by \be H_\alpha=H_\alpha^0+\lambda_\alpha H_\alpha^I,\qquad H_\alpha^0=\omega_\alpha^a a_\alpha^\dagger a_\alpha+\omega_\alpha^b
b_\alpha^\dagger b_\alpha, \quad H_\alpha^I=a_\alpha^\dagger b_\alpha+b_\alpha^\dagger a_\alpha. \label{ex21} \en It is natural to interpret
the mean values of the operators $\hat n^{(a)}_\alpha$ and $\hat n^{(b)}_\alpha$ as {\em local density operators}  of the two populations in
the cell $\alpha$: if the mean value of, say, $\hat n^{(a)}_\alpha$, in the state of the system is equal to one, this means that the density of
$\Sc_1$ in the cell $\alpha$ is very high. Notice that $H_\alpha=H_\alpha^\dagger$, since all the parameters, which in general are assumed to
be cell--depending (to allow for the description of an anisotropic situation), are real and positive numbers. The CAR are
 \be \{a_\alpha,a_\beta^\dagger\}=\{b_\alpha,b_\beta^\dagger\}=\delta_{\alpha,\beta}\,\1, \qquad
\{a_\alpha^\sharp,b_\beta^\sharp\}=0. \label{ex22} \en Of course, the full hamiltonian $H$ must consist of a sum of all the different
$H_\alpha$ plus another contribution, $H_{diff}$, responsible for the diffusion of the populations all around the lattice. A natural choice for
$H_{diff}$, in view of the above rules, is the following: \be H_{diff}=\sum_{\alpha,\beta}p_{\alpha,\beta}\left\{\gamma_a\left(a_\alpha
a_\beta^\dagger+a_\beta a_\alpha^\dagger\right)+\gamma_b\left(b_\alpha b_\beta^\dagger+b_\beta b_\alpha^\dagger\right)\right\}, \label{ex23}
\en where also $\gamma_a$, $\gamma_b$ and the $p_{\alpha,\beta}$ are real quantities. In particular, $p_{\alpha,\beta}$ can only be 0 or 1
depending on the possibility of the populations to move from cell $\alpha$ to cell $\beta$ or vice-versa. For this reason they are considered
as {\em diffusion coefficients}. Notice that a similar role is also played by $\gamma_a$ and $\gamma_b$. $H=\sum_{\alpha}H_\alpha+H_{diff}$
obeys the three rules of Section II. Indeed, if there is no interaction between $\Sc_1$ and $\Sc_2$, and between members of the same species
localized in different cells of $\R$, it is easy to check that the densities of $\Sc_1$ and $\Sc_2$ stay constant in all the cells: $\hat
n_\alpha^{(a)}(t)=\hat n_\alpha^{(a)}(0)$ and $\hat n_\alpha^{(b)}(t)=\hat n_\alpha^{(b)}(0)$, for all $\alpha$. Hence {\bf Rule 2} holds true.
{\bf Rule 3} is also satisfied, since $\Sc_1$ and $\Sc_2$ cannot move outside $\R$: it is again possible to find an operator, related to the
total number of members of $\Sc$ distributed all along $\R$, which commutes with $H$, so that this {\em global density} stays constant in time.
Concerning {\bf Rule 1}, we see that this is applied several times in the definition of $H$. For instance we have the contribution
$a_\alpha^\dagger b_\alpha+b_\alpha^\dagger a_\alpha$, which shows how {\bf Rule 1} is applied in the interaction between $\Sc_1$ and $\Sc_2$
in the cell $\alpha$, but we also have $a_\alpha a_\beta^\dagger+a_\beta a_\alpha^\dagger$, which is again {\bf Rule 1}, but applied to $\Sc_1$
in different cells. And so on. Again, we refer to \cite{ff3} for the analysis of the equations of motion arising from this hamiltonian.

\subsection{Last example: stock market}

In recent years we have proposed several hamiltonians describing simplified stock markets, \cite{bag1}-\cite{bag4}. The one we discuss here,
the {\em most efficient} proposal, so far, was first introduced in \cite{bag4}.

Let us consider $N$ different traders $\tau_1$, $\tau_2$, $\ldots$, $\tau_N$, exchanging $L$ different kind of shares $\sigma_1$, $\sigma_2$,
$\ldots$, $\sigma_L$. Each trader has a starting amount of cash, which is used during the trading procedure: the cash of the trader who sells a
share increases while the cash of the trader who buys that share consequently decreases. The absolute value of these variations is the price of
the share at the time in which the transaction takes place. It is clear that the above-mentioned division of $\Sc$ in just two plus one
components, $\Sc_1$, $\Sc_2$ and $\M$, must be extended here, while the main ideas are unchanged. It is convenient to introduce a set of
bosonic operators which are listed, together with their economical meaning, in the following table. We are adopting here latin indexes to label
the traders and greek indexes for the shares: $j=1,2,\ldots,N$ and $\alpha=1,2,\ldots,L$.

\vspace{5mm} {\hspace{-.8cm}
\begin{tabular}{|c||c||c||c||c|} \hline    &the operator and.. &...its economical meaning               \\
\hline  & $a_{j,\alpha}$  & annihilates a share $\sigma_\alpha$  in the portfolio of $\tau_j$    \\
\hline   & $a_{j,\alpha}^\dagger$ & creates a share $\sigma_\alpha$  in the portfolio of $\tau_j$   \\
\hline  & $\hat n_{j,\alpha}=a_{j,\alpha}^\dagger a_{j,\alpha}$ &counts the number of share $\sigma_\alpha$  in the portfolio of $\tau_j$   \\
\hline\hline
\hline  & $c_j$  & annihilates a monetary unit in the portfolio of $\tau_j$    \\
\hline   & $c_j^\dagger$ & creates a monetary unit in the portfolio of $\tau_j$   \\
\hline  & $\hat k_j=c_j^\dagger c_j$ &counts the number of monetary units in the portfolio of $\tau_j$   \\
\hline\hline
\hline  & $p_\alpha$  &  lowers the price of the share $\sigma_\alpha$ of one unit of cash   \\
\hline   & $p_\alpha^\dagger$ &  increases the price of the share $\sigma_\alpha$ of one unit of cash   \\
\hline  & $\hat P_\alpha=p_\alpha^\dagger p_\alpha$ &gives  the  value of the share $\sigma_\alpha$  \\
\hline\hline

\end{tabular}

\vspace{4mm}

\indent Table 1.-- List of operators and of their {\em economical} meaning.}

\vspace{3mm}

These operators are bosonic in the sense that they satisfy the following commutation rules \be [c_j,c_k^\dagger]=\1\,\delta_{j,k},\quad
[p_\alpha,p_\beta^\dagger]=\1\,\delta_{\alpha,\beta}\quad
[a_{j,\alpha},a_{k,\beta}^\dagger]=\1\,\delta_{j,k}\delta_{\alpha,\beta},\label{ex31}\en while all the other commutators are zero.
 We assume that the hamiltonian of the market,  $\hat H$, can be written as $\hat H=H+H_{prices}$, where
 \be
\left\{
\begin{array}{ll}
H=H_0+ \lambda\,H_I, \mbox{ with }  \\
H_0 = \sum_{j,\alpha}\,\omega_{j,\alpha}\, \hat n_{j,\alpha}+\sum_{j}\,\omega_j\, \hat k_j\\
 H_I = \sum_{i,j,\alpha}\,p_{i,j}^{(\alpha)}\left(a_{i,\alpha}^\dagger a_{j,\alpha}c_i^{\hat P_\alpha} {c_j^\dagger}^{\hat P_\alpha}+h.c.\right). \\
\end{array}
\right. \label{ex32} \en Here h.c. stands for hermitian conjugate, $c_i^{\hat P_\alpha}$ and ${c_j^\dagger}^{\hat P_\alpha}$ are defined as in
\cite{bag2}, and $\omega_{j,\alpha}$,  $\omega_j$ and $p_{i,j}^{(\alpha)}$ are positive real numbers. In particular these last coefficients
assume different values depending on the possibility of $\tau_i$ to interact with $\tau_j$ and exchanging a share $\sigma_\alpha$: for instance
$p_{2,5}^{(1)}=0$ if there is no way for $\tau_2$ and $\tau_5$ to exchange a share $\sigma_1$. It is natural to put $p_{i,i}^{(\alpha)}=0$ and
$p_{i,j}^{(\alpha)}=p_{j,i}^{(\alpha)}$.

Going back to (\ref{ex32}), we observe that $H$ obeys {\bf Rule 2} of Section 2, since, if  there is no interaction between the traders, then
$\lambda=0$ and, as a consequence, $H=H_0$: $[H_0,\hat n_{j,\alpha}]=0$, for all $j$ and $\alpha$. As for $H_I$, this is written obeying {\bf
Rule 1}:
 the action of a single contribution of $H_I$, $a_{i,\alpha}^\dagger a_{j,\alpha}c_i^{\hat P_\alpha}
{c_j^\dagger}^{\hat P_\alpha}$, on a vector number which extends those introduced in Section II,
$\varphi_{\{n_{j,\alpha}\};\{k_j\};\{P_\alpha\}}$, is proportional to another vector $\varphi_{\{n_{j,\alpha}'\};\{k_j'\};\{P_\alpha'\}}$ with
just 4 different quantum numbers. In particular $n_{j,\alpha}$, $n_{i,\alpha}$, $k_j$ and $k_i$ are replaced respectively by $n_{j,\alpha}-1$,
$n_{i,\alpha}+1$, $k_j+P_\alpha$ and $k_i-P_\alpha$ (if this is larger or equal than zero, otherwise the vector is annihilated). This means
that $\tau_j$ is selling a share $\sigma_\alpha$ to $\tau_i$ and earning money from this operation. For this reason it is convenient to
introduce the following {\em selling} and {\em buying} operators: \be x_{j,\alpha}:=a_{j,\alpha}\,{c_j^\dagger}^{\hat P_\alpha},\qquad
x_{j,\alpha}^\dagger:=a_{j,\alpha}^\dagger\,{c_j}^{\hat P_\alpha} \label{ex34}\en With these definitions and using the properties of the
coefficients $p_{i,j}^{(\alpha)}$ we can rewrite $H_I$ as \be
 H_I = 2\,\sum_{i,j,\alpha}\,p_{i,j}^{(\alpha)}x_{i,\alpha}^\dagger\,x_{j,\alpha} \Rightarrow H=\sum_{j,\alpha}\,\omega_{j,\alpha}\, \hat n_{j,\alpha}+\sum_{j}\,\omega_j\, \hat k_j+2\,\lambda\,\sum_{i,j,\alpha}\,p_{i,j}^{(\alpha)}x_{i,\alpha}^\dagger\,x_{j,\alpha}
\label{ex35}\en The role of  $H_{price}$ in \cite{bag4} was to fix the time evolution of the operators $\hat P_\alpha$, $\alpha=1,2,\ldots,L$.
 The
hamiltonian $\hat H$ corresponds to a closed market where the money and the total number of shares of each type are conserved. Indeed, calling
$\hat N_\alpha:=\sum_{l=1}^N\hat n_{l,\alpha}$ and $\hat K:=\sum_{l=1}^N\hat k_{l}$ we see that, for all $\alpha$, $[\hat H,\hat
N_\alpha]=[\hat H,\hat K]=0$. Hence $\hat N_\alpha$ and $\hat K$ are integrals of motion, as expected: {\bf Rule 3} is satisfied. We refer to
\cite{bag4} for the analysis of the time evolution of  the {\em portfolio operator} of the trader $\tau_l$, $ \hat
\Pi_l(t)=\sum_{\alpha=1}^L\hat P_\alpha(t)\,\hat n_{l,\alpha}(t)+\hat k_l(t)$.

\section{Further considerations and conclusions}

The same rules have already been adopted for systems which are not closed, i.e. for those systems which exchange {\em something} with the
environment. This is discussed, for instance, in \cite{bag5}: again, the main idea is that we can use  creation and annihilation operators also
in the description of the reservoir, and in modeling an exchange between the system and the reservoir. This exchange is described adding in the
hamiltonian a contribution obeying {\bf Rule 1}, while {\bf Rule 2} has to be intended here in the following way: if $\Sc$ does not interact
with the reservoir, no decay is allowed. {\bf Rule 3} is recovered for some global quantity which mixes the degrees of freedom of the reservoir
and of the system. The conclusion is, therefore, that our rules can be used also in more general, and sometimes more useful, contexts.



\addcontentsline{toc}{section}{\refname}
\begin{thebibliography}{9}




\bibitem{bag1} F. Bagarello, {\em An operatorial approach to stock markets},
  J. Phys. A, {\bf 39}, 6823-6840 (2006)

\bibitem{bag2} F. Bagarello, {\em Stock Markets and Quantum Dynamics: A Second Quantized Description},
 Physica A, {\bf 386}, 283-302 (2007)

\bibitem{bag3} F. Bagarello, {\em Simplified Stock markets and their quantum-like
dynamics},  Rep. on Math. Phys., {\bf 63}, nr. 3, 381-398 (2009)

\bibitem{bag4} F. Bagarello, {\em A quantum statistical approach to simplified stock markets}.
Physica A, \textbf{388}, 4397--4406, 2009.


\bibitem{ff1} F. Bagarello, F. Oliveri, {\em Quantum Modeling of Love Affairs}. Proceedings Wascom 2009,  A. M. Greco, S. Rionero, T. Ruggeri eds.,  7--14, World Scientific, Singapore, 2010.

\bibitem{ff2} F. Bagarello, F. Oliveri, {\em An operator--like description of love affairs}, SIAM J.
Appl. Math.,  {\bf 70}, 3235--3251, 2011.

\bibitem{bag5} F. Bagarello, {\em Damping in quantum love affairs},  Physica A, {\bf 390}, 2803--2811, 2011.

\bibitem{ff3} F. Bagarello, F. Oliveri, {\em An operator description of interactions between populations with applications to migration},
 Math. Mod. and Meth. in Appl. Sci., submitted


\bibitem{baa} B.E. Baaquie, {\em Quantum Finance}, Cambridge
University Press, 2004

\bibitem{chou} O. Al. Choustova, {\em Quantum Bohmian model for financial market}, Physica A, {\bf 374}, 304--314, 2007.

\bibitem{haven} E. Haven, {\em Pilot-wave theory and financial option pricing},
Int. Jour. Theor. Phys., {\bf 44}, No. 11, 1957-1962, 2010

\bibitem{jimenez} E. Jimenez, D. Moya, {\em Econophysics: from game theory and information theory to quantum mechanics}, Physica A, {\bf 348}, 505--543, 2005.


\bibitem{khre} A. Khrennikov, {\em Ubiquitous quantum structure: from psychology to finances}, Springer,
Berlin, 2010.

\bibitem{tulu} S. I. Melnyk, I. G. Tuluzov, {\em Quantum analog of the Black-Scholes formula (market of financial derivatives as a continuous weak measurement)}, Elect. Journ. Theor. Phys., {\bf 5}, No 18, 95-108, 2008

\bibitem{seg} W. Segal, I. E. Segal, {\em The Black–Scholes pricing formula in the quantum context}, Proc. Natl. Acad. Sci. USA, {\bf 95},  4072–4075, 1998


\bibitem{mer} E. Merzbacher, {\em Quantum Mechanics}, Wiley, New York, 1970.





























\end{thebibliography}
\end{document}